\documentclass[prb,twocolumn] {revtex4-1} 
\usepackage{graphicx}
\usepackage{url}
\usepackage{upgreek}
\usepackage{amsmath}
\newcommand{\mymu}{{\ensuremath \upmu}}
\begin{document}
\title{A determination of the local acceleration of gravity for the NIST-4 watt balance}
\author{E.J.~Leaman} \email{Present address: Virginia Polytechnic Institute, Blacksburg, VA, USA}
\author{D.~Haddad}
\author{F.~Seifert}
\author{L.S.~Chao}
\author{A.~Cao}
\author{J.R.~Pratt}
\author{S.~Schlamminger}
\author{D.B. Newell}  \email{Email: David.Newell@nist.gov}

\affiliation{Fundamental Electrical Measurements Group, National Institute of Standards and Technology, Gaithersburg, MD 20899}


\date{\today}

\begin{keywords}
{fundamental constants, Planck constant, gravity, international system of units, watt balance}
\end{keywords}

\begin{abstract}
A new watt balance is being constructed at the National Institute of Standards and Technology (NIST) in preparation for the redefinition of the International System of Units and the realization of mass through an exact value of the Planck constant. The total relative uncertainty goal for this instrument of a few parts in $10^{8}$ requires that the local acceleration due to gravity be known at the location of a test mass with a relative uncertainty on the order of only a few parts in $10^{9}$. To make this determination, both the horizontal and vertical gradients of gravity must be known such that gravity may be tied from an absolute reference in the laboratory to the precise mass location. We describe the procedures used to model and measure gravity variations throughout the laboratory and give our results.
\end{abstract}
\maketitle

\section{Introduction}
Resolution 1 of the 24th Meeting of the General Conference  on Weights and Measures (CGPM) outlines a revised International System of Units (SI) based upon fixed numerical values of invariants of nature~\cite{CGPM}. A consequence is that the unit of mass will be realized through an assigned value of the Planck constant, $h$. A method for realizing mass via $h$ utilizes a watt balance that requires the determination of the local acceleration of gravity, $g$, with parts in $10^{9}$ accuracy. At the National Institute of Standards and Technology (NIST), significant progress has been made with the existing  watt balance (NIST-3) towards providing the necessary experimental results~\cite{nist3} for the possible SI redefinition to occur at the 26th meeting of the CGPM. In preparation for disseminating mass in the revised SI, a new robust NIST watt balance, NIST-4, is being constructed~\cite{nist4}. We discuss our methods for modeling the new NIST-4 laboratory and major instrument components, our measurements of gravity and final results, and our uncertainty analysis~\cite{nist4grav}.

\section{Watt Balances and the Local Acceleration Due to Gravity}
To summarize the operational principle of a watt balance, the measurements involved relate the Planck constant to mass, length, and frequency through 
\begin{equation}
m =\frac{hC}{gv},
\label{eq:prin}
\end{equation}
where $m$ is mass, $g$ is the local acceleration due to gravity, $v$ is a velocity, and $C$ is a combination of frequencies and scalar constants. With a total uncertainty goal for the watt balance on the order of a few parts in $10^{8}$, equation~\ref{eq:prin} highlights the importance of determining $g$ at the location of the mass standard used in a watt balance to an accuracy of a few parts in $10^{9}$ or lower such that its uncertainty is negligible in the final watt balance result.

A challenge in evaluation of $g$ for the recent NIST-3 result~\cite{nist3grav} was that the apparatus was physically in the way. Moreover, the measurements of the absolute value of $g$ were performed intermittently, with corrections being applied for
tidal, polar motion, and atmospheric effects during the operation of the watt balance.  The determination of gravity for NIST-4 follows that of more thorough evaluations of gravity of other watt balance efforts~\cite{lnegrav,metasgrav,icac,NRCgrav}. Ties (the difference in gravity between two points) were measured between absolute measurement locations and the horizontal-plane position of the NIST-4 test mass along the vertical concentric symmetry axis of the permanent magnet used to generate the magnetic field for NIST-4. This was done before the mass pan and related components were installed, which will prevent such measurements in the future. These measurements allow us to validate and refine the gravity model of the room and the apparatus  so that the model may be used for gravity determinations in the future.  Furthermore, a permanent relative gravimeter will be installed in the location of previously-recorded measurements of absolute $g$ for real time monitoring.

To stay consistent with the field of absolute gravimetry, the non SI unit of acceleration, the gal (symbol Gal), is used in this text. It is defined as 1\,Gal = 1\,cm/s$^{2}$~\cite{sibro}.

\section{Modeling}

\begin{figure}[h!]
\includegraphics[width=3.3in]{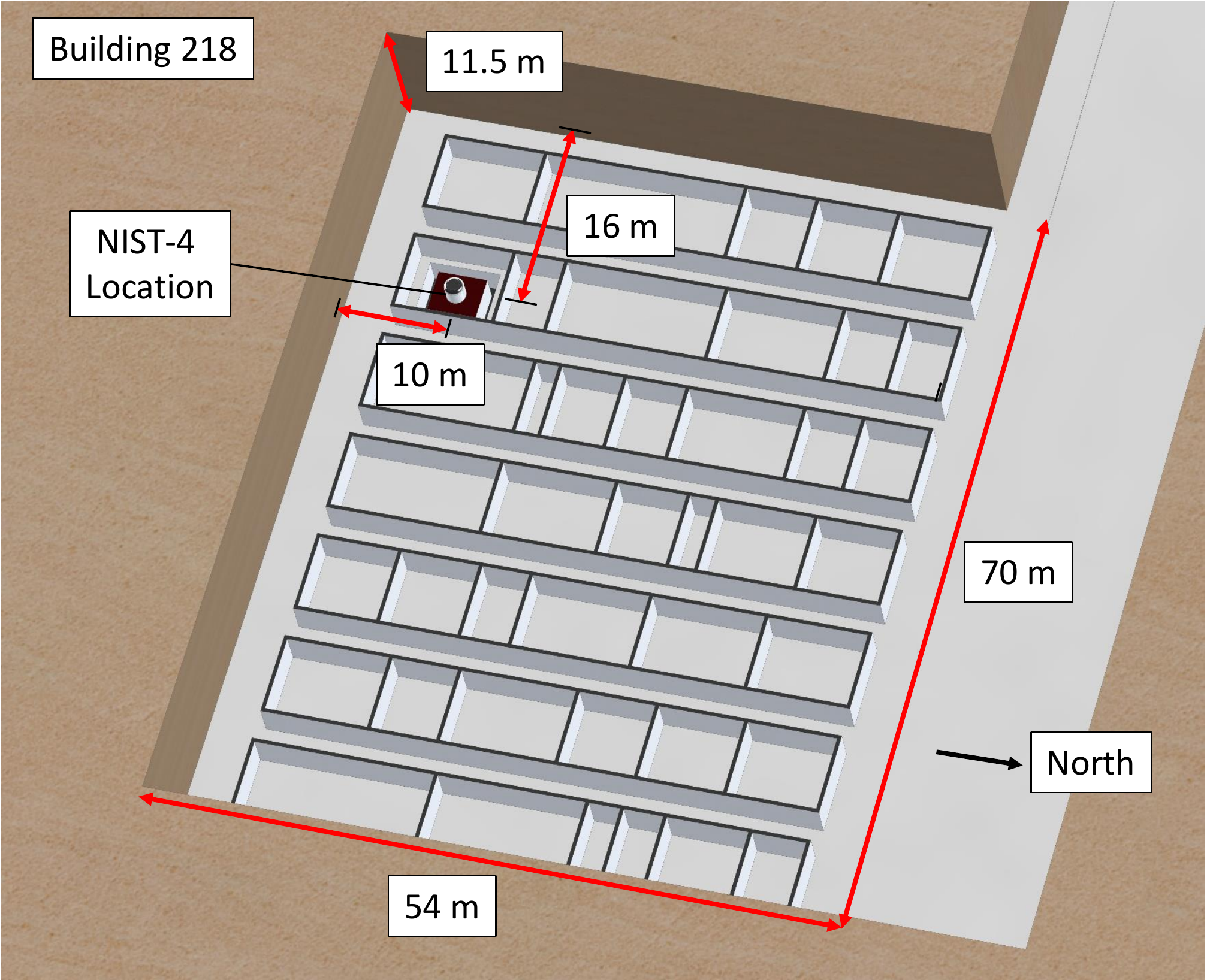}
\label{fig:218}
\end{figure}

\begin{figure*}[t!]
\centering
\includegraphics[width=7in]{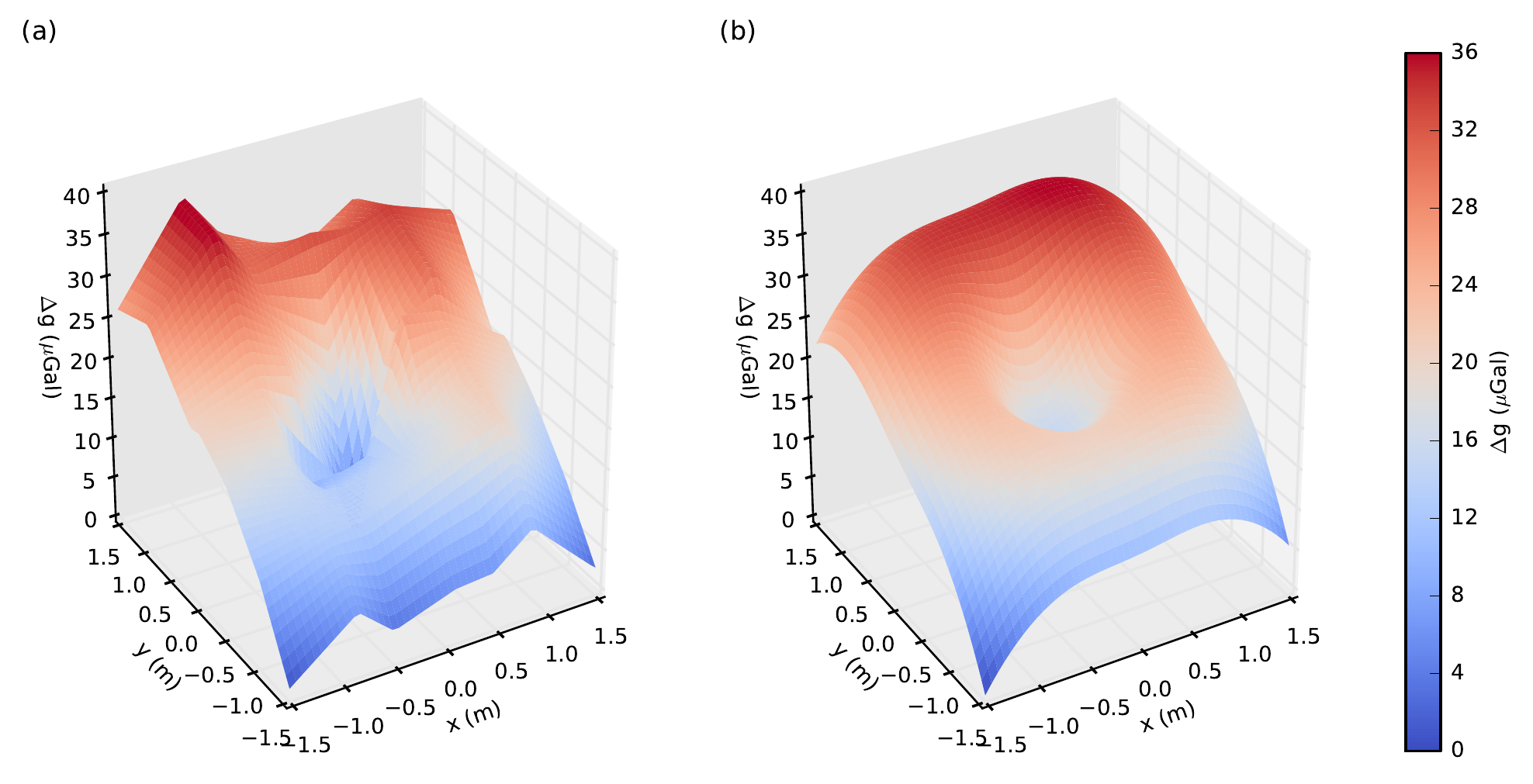}
\caption{(a) Mapping of gravity across the concrete slab at a height of 25.9\,cm (interpolated between data points) and (b) model of the subterranean laboratory and concrete slab at the same height. 1\,$\mymu$Gal = $10^{-8}$\,m/s$^{2}$}
\label{fig:gmaps}
\end{figure*}

The model of the NIST-4 laboratory and components was created using the method of right rectangular prisms~\cite{nagy66}. Following~\cite{NRCgrav}, this method was chosen because it provides a simple analytical solution for the gravitational attraction in the vertical direction at any point in space. The accuracy of the solution is then only dependent on the accuracy in the gravitational constant, $G$, and object densities and dimensions. The coordinate system was chosen such that most significant rectangular components were aligned, and non-rectangular components were approximated with simple algorithms to break them into a finite number of prisms of varying dimensions. 

The laboratory for the NIST-4 watt balance is in a subterranean building approximately 11.5\,m underground and is much closer to the building limits on the south and west facing walls than on the others, as shown in Figure~\ref{fig:218}. This placement results in a significant change in gravity as a function of horizontal position  across the room. The watt balance sits on the center of a 60\,metric ton concrete slab (dimensions: 4\,m$\times$4\,m$\times$2.4\,m) , with the magnet and horizontal position of the test mass offset by approximate 25\,cm in both the $x$ and $y$ directions. 

\begin{figure}[h]
\centering
\includegraphics[width=3.3in]{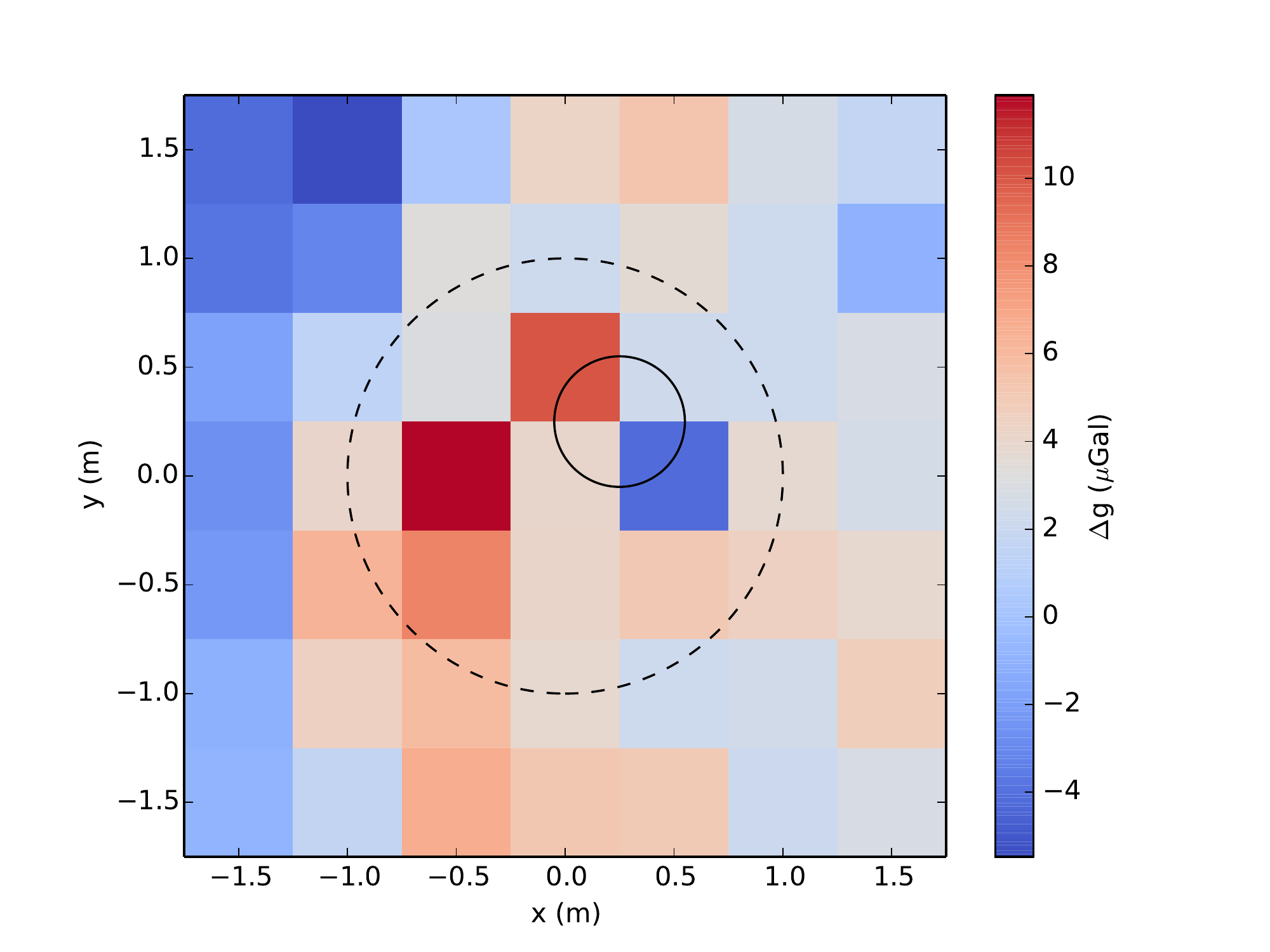}
\caption{Difference between model and measurements of gravity changes throughout NIST-4 laboratory in July 2013. The color coded residuals are plotted near the points, where the measurements were made, i.e., a 7 by 7 grid with a 0.5\,m spacing.  At the time of the measurement the laboratory was void of equipment. The dashed circle indicates the hole in the concrete block, that was filled with sand in December 2013. The solid circle indicates the position, where the magnet was placed subsequently.}
\label{fig:residuals}
\end{figure}

The greatest contributors to changes in gravity throughout the laboratory are the concrete slab and the surrounding earth, which were assumed to have average densities of 2.4\,g/cm$^{3}$ and 1.4\,g/cm$^{3}$, respectively. A horizontal mapping of the gravity at a height of 25.9\,cm from the  top surface of the concrete slab was made in July 2013 using a Scintrex CG-5 Autograv\footnote{Certain commercial equipment, instruments, or materials are identified in this paper in order to specify the experimental procedure adequately. Such identification is not intended to imply recommendation or endorsement by the National Institute of Standards and Technology, nor is it intended to imply that the materials or equipment identified are necessarily the best available for the purpose} relative gravimeter~\cite{scintrex}. At that time, a 1\,m diameter hole existed in the center of the concrete slab, extending through the bottom. The results were used as an initial confirmation of the model, which neglects the building foundation and other relatively insignificant changes in density. The model extended the earth below and on the sides of the laboratory to approximately 100\,m from the center of the slab. Figure~\ref{fig:gmaps} shows both the measurements and the model at the same height. Note that the reduction in gravitational attraction due to the hole in the slab appears off center in the measurements plot. This is believed to be measurement error due to using a coarse measurement grid and because only two measurements were used for an average in each position (data was smoothed by interpolating between measurements). Figure~\ref{fig:residuals} shows the difference between the model and the measurements at the grid spacing of the measurement. Generally, the model is in good agreement with the measurement. The model reproduces the gradient along the $y$-direction. The largest residuals occur at the position of the hole and can be explained with worse measurement conditions. The hole was covered with plywood, providing a less stable base than the block itself. 

\begin{figure}[h]
\centering
\includegraphics[width=3.3in]{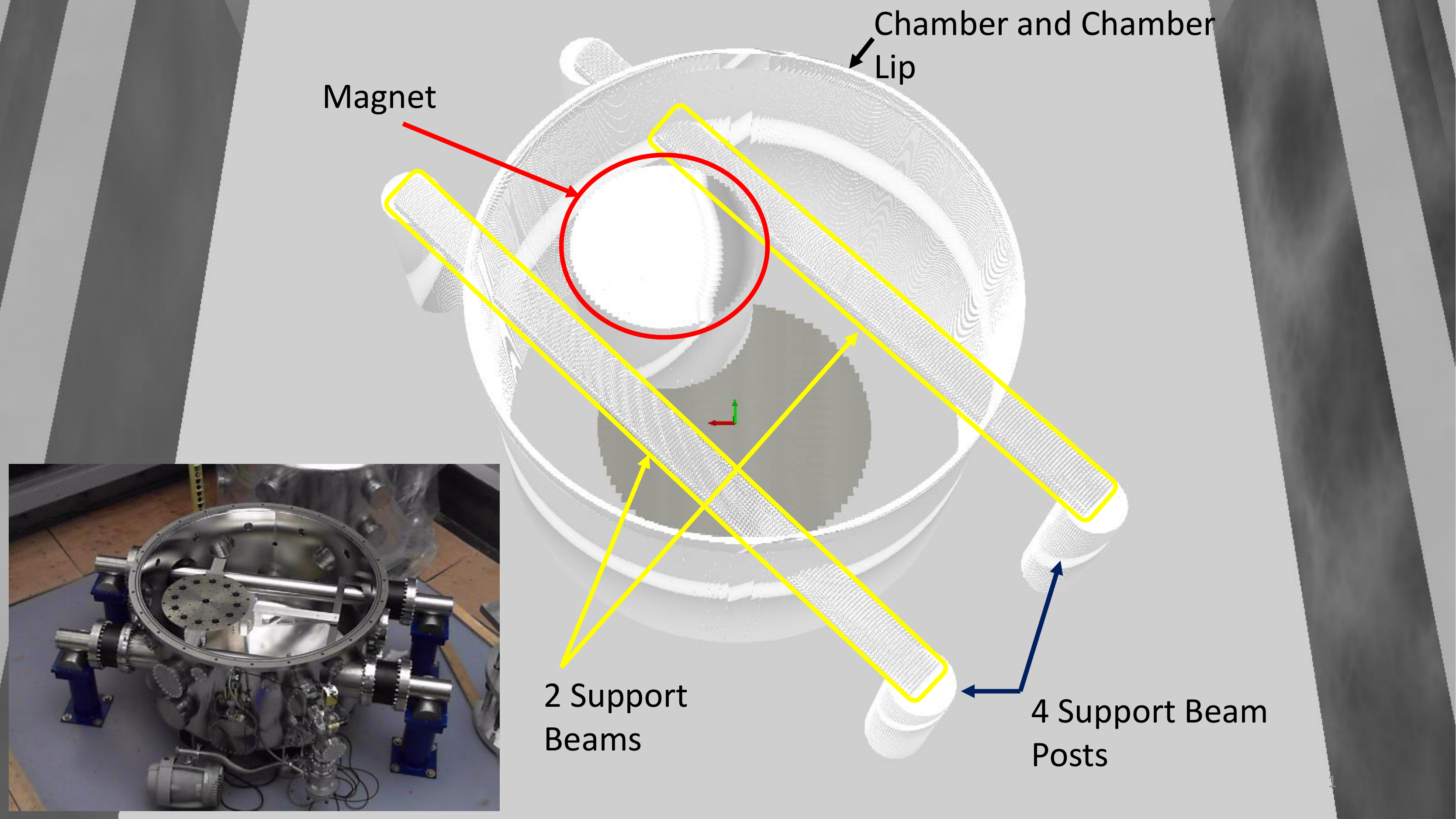}
\caption{Visualization of modeled components of the NIST-4 watt balance}
\label{fig:comps}
\end{figure}

\begin{figure}[h]
\centering
\includegraphics[width=3.3in]{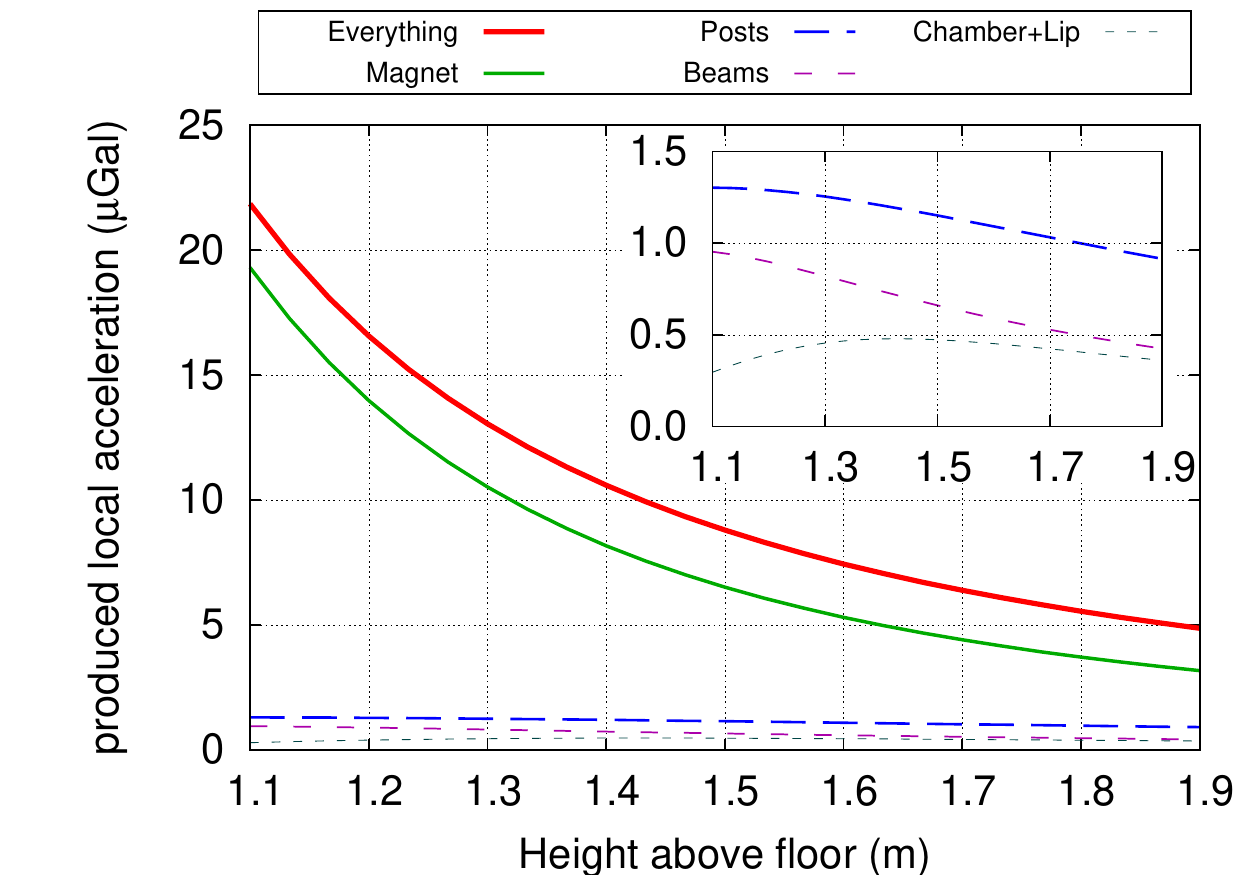}
\caption{Modeled gravitational attraction of massive NIST-4 components. For reference, the top surface of the magnet is about 0.85\,m from the floor.}
\label{fig:contributions}
\end{figure}

\begin{figure}[h]
\centering
\includegraphics[width=3.3in]{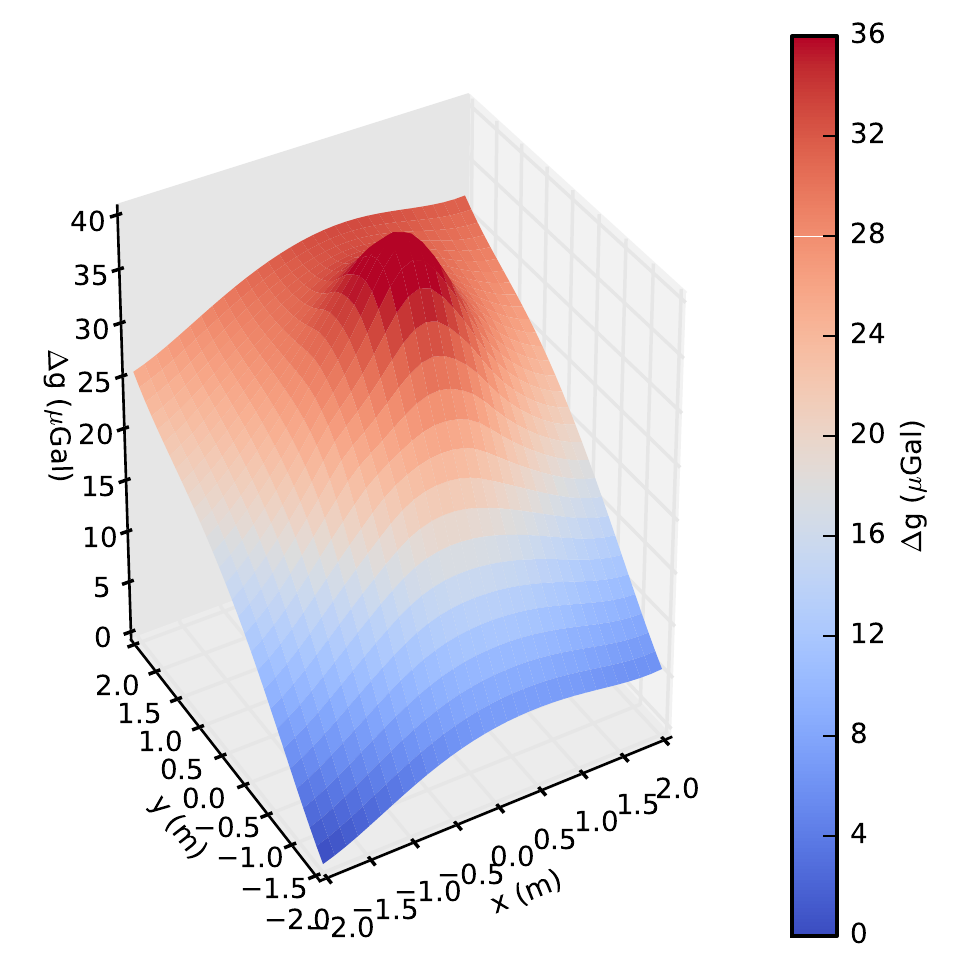}
\caption{Full model of gravity in NIST-4 lab at a height of 130\,cm.}
\label{fig:fmodel}
\end{figure}

In order to use the model to predict gravity variations between absolute reference points and the test mass location, several of the more massive components of NIST-4 were modeled to account for instrument self-attraction. A visualization of these modeled components is shown in Figure~\ref{fig:comps}, along with a photograph of the lower half of the vacuum chamber during installation. The approximate masses of these objects is given in Table~\ref{table:mass}. The modeled gravitational attraction of each of these components as a function of height (on axis with the center of the magnet) is shown in Figure~\ref{fig:contributions}. Note that the magnet is much heavier than any of the other components and thus is responsible for most of the instrument self-attraction. The other modeled components create a gravitational attraction on the order of about $10^{-9}$ each. Thus, no other components were modeled as their effect was determined to be negligible. The full model, including major NIST-4 components, at a height of 130\,cm (approximately the normal height of the test mass) is shown in Figure~\ref{fig:fmodel}. 
The effect of the vacuum bell jar was modeled and the gravitational attraction was found to be below 1\,$\mu$Gal.

\begin{table}[h]
\centering
\caption{Masses of components in the computer model to calculate the gravitational field.}
\begin{tabular}{lr}
\hline
Component & Mass (kg) \\
\hline
Magnet & 816 \\
Bottom Half of Chamber and Lip & 127 \\
One Support Beam & 216 \\
One Support Beam Post  & 42 \\
\hline
\end{tabular}
\label{table:mass}
\end{table}

\section{Measuring Gravity}
In this section, we discuss our measurements of gravity and further model validation for our final determination of absolute gravity at the mass position.

\subsection{Vertical gravity gradients (VGGs)}
\label{sec:vgg}
\begin{figure}[h]
\centering
\includegraphics[width=3.3in]{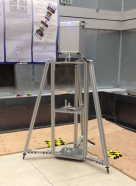}
\caption{The stand used to measure vertical gravity gradients with a CG-5 relative gravimeter.}
\label{fig:CG5stand}
\end{figure}

\begin{figure}[h]
\centering
\includegraphics[width=3.3in]{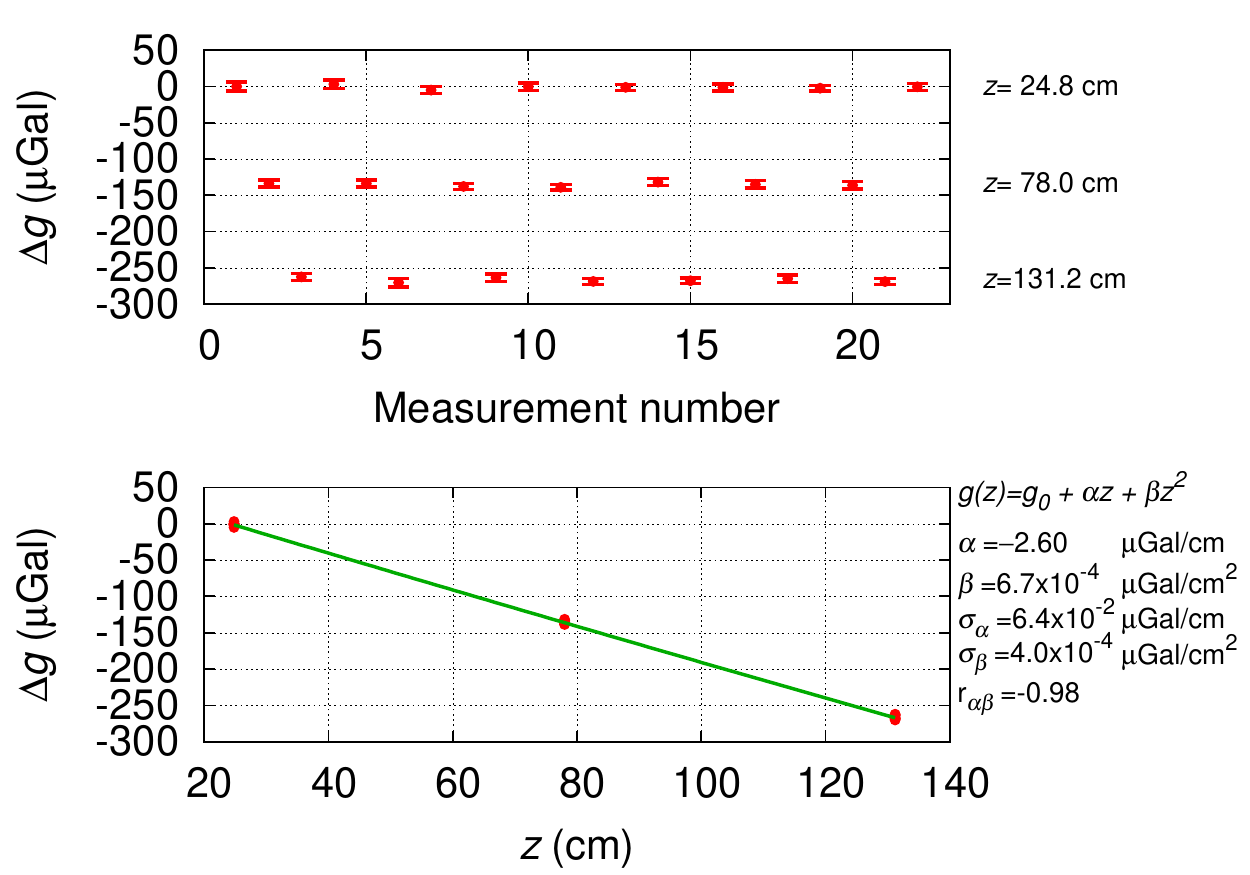}
\caption{Example of a measurement of the vertical gravity gradient at the north-east corner. The top graph shows the raw data measured by the relative gravimeter at three vertical positions. The values have been drift corrected and an offset has been subtracted. The lower graph shows the measurement versus vertical position. The results of the least squares adjustment of the second order polynomial (solid line) are shown to the right of the graph.}
\label{fig:vgg}
\end{figure}

In order to make tie measurements more versatile and allow for long-term future use, the vertical gravity gradients (VGGs) (as a function of height from the top surface of the concrete slab) were measured at the north-east (NE), south-east (SE), and south-west (SW) corners of the lab (where tie measurements would be made). The VGG above the NIST-4 magnet was also measured. This allows for the vertical translation of gravity values at each point where the VGG is measured and  tie measurements at the different heights as absolute measurements. The VGG determinations were made by taking CG-5 measurements at heights of approximately 25\,cm, 78\,cm, and 130\,cm from the top surface of the concrete slab, each integrated over 2\,min. The stand used for these measurements is shown in Figure~\ref{fig:CG5stand}. The first reading was taken at the lowest height, the second at the middle height, and the third at the highest. This process was repeated five times for six readings at each height and ending with a seventh reading on the lowest height so that instrument drift may be removed. An example of one such measurement is shown in Figure~\ref{fig:vgg}.  VGGs were measured above the magnet in a similar fashion, except heights of about 115\,cm, 120\,cm, 125\,cm, and 130\,cm from the floor were used (the top surface of the magnet is about 85\,cm from the floor), and only two total runs were made. For this measurement, the CG-5 was supported by carefully measured and leveled optic stands. 

Using a least-squares analysis, the VGG data was  fitted to the second-order polynomial 
\begin{equation}
g(z)=\beta z^2 + \alpha z + g_0,
\label{eq:g(z)}
\end{equation}
where $z$ is the vertical distance from the reference and $g_0$ is the value $g(z=0)$. The change in gravity for a vertical translation from heights $z_1$ to $z_2$ is then
\begin{equation}
v = g(z_2)-g(z_1) = \beta (z_2^2 - z_1^2) + \alpha (z_2 - z_1).
\end{equation}
Accounting for correlation between the coefficients $\alpha$ and $\beta$, the variance in the vertical translation is:
\begin{eqnarray}
\sigma_{v}^2 &=& \sigma_{\beta}^2 (z_2^2 - z_1^2)^2 + \sigma_{\alpha}^2 (z_2 - z_1)^2 \nonumber \\
			& &  		  + 2 (z_2^2 - z_1^2) (z_2 - z_1) \sigma_{\alpha \beta}^2.
\end{eqnarray}
where $\sigma_{\beta}^2$ and $\sigma_{\alpha}^2$ are the variances in $\beta$ and $\alpha$, respectively, and $\sigma_{\alpha \beta}^2=r_{\alpha\beta}\sigma_\alpha\sigma_\beta$ is the covariance between $\alpha$ and $\beta$ with $r_{\alpha\beta}$ being their correlation coefficient. 

\subsection{Tie Measurements}
\begin{figure}[h] 
\centering
\includegraphics[width=3.3in]{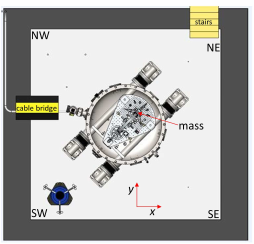}
\caption{An aerial view of NIST-4 and the NIST-4 laboratory. The abbreviations in the four corners denote the geographic directions (NW = north-west, NE = north-east SE = south-east, SW = south-west). The device drawn in the south-west corner is a superconducting relative gravimeter, that will be installed in the fall of 2014 and was not at the site when the measurements reported here were performed.}
\label{fig:lab}
\end{figure}

A series of measurements were made using the CG-5 in order to determine the change in gravity between locations in the lab and ultimately tie the mass location to the southwest and southeast corners of the room, where absolute measurements were made in the summer of 2013 using a Micro-g LaCoste FG5 free fall gravimeter~\cite{microg}. An aerial view of the NIST-4 laboratory is shown in Figure~\ref{fig:lab}. The measurements were made by moving the CG-5 between two locations, taking at least six measurements at each. Note, however, that all absolute readings were made at a height of 130\,cm from the floor, while the CG-5 was used at various heights. This necessitates a vertical translation when tying to the mass position as discussed in Section~\ref{sec:vgg}. The ties were created by taking the difference between the average reading at each location, and the variance in the tie was found by summing in quadrature the standard deviation of the mean in each location. This process was repeated five times to establish three ties from SW and one from SE to the mass position (M). The fourth tie was between SW and NE. The first SW-to-magnet tie was made at a SW height of only 25.9\,cm, whereas the second was made at about 131\,cm to reduce the amount of vertical translation needed. 

\section{Determining Absolute Gravity at the Mass Position}
\label{sec:tie}
\begin{figure}[h]
\centering
\includegraphics[width=3.3in]{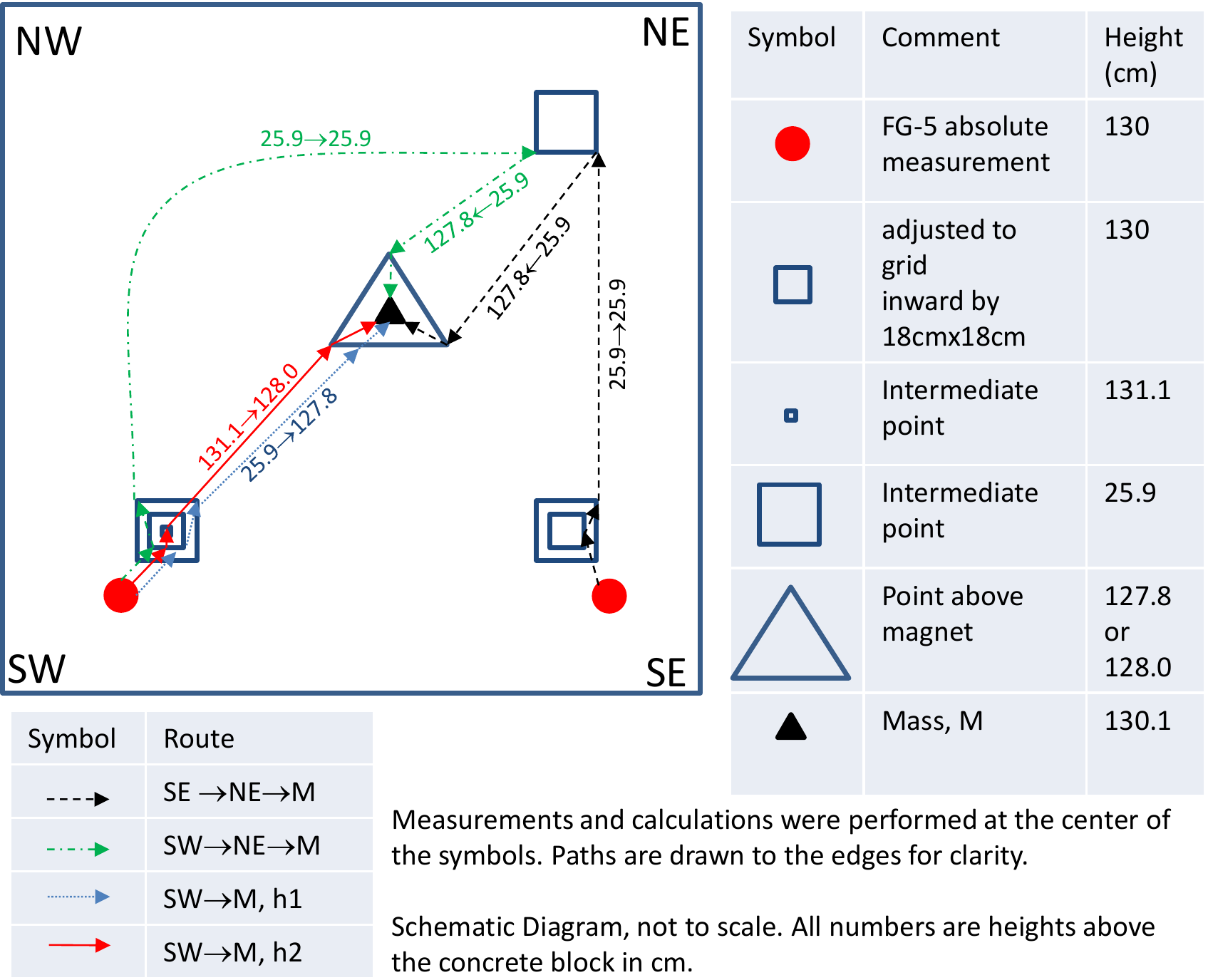}
\caption{Diagram showing the four unique ways to determine absolute $g$ at the mass position with the measurements made.}
\label{fig:ties}
\end{figure}

The tie measurements give four unique paths to tie the absolute gravity determinations at SE and SW to the mass position, shown schematically in Figure~\ref{fig:ties}. Each absolute measurement had to first be shifted towards the center of the concrete slab 18\,cm$\times$18\,cm to account for a change in reference points between July 2013, when the absolute measurements were made, and the summer of 2014. This was done using only the model and resulted in a small change in $g$ of about 2\,$\mymu$Gal at both SE and SW. The shift was assigned an uncertainty of half its value. Next, the VGG measurements were used to translate $g(z=130\,\mbox{cm})$ to the height where a tie was made (in most cases 25.9\,cm). The result could then be directly tied to NE from either SE or SW and then to M (making two possible paths), or from SW to M using one of the two ties made directly above the magnet. Finally, this tied value of $g$ above the magnet can then be translated vertically to the precise position of the NIST-4 test mass. The numerical values of each step in each of the four ties are given in Table~\ref{table:ties}. The uncertainties in the final two digits of each nominal value are enclosed in parenthesis following the value.

\begin{table*}[t!]
\centering
\caption{Four paths tying absolute gravity measurements to the position of the test mass}
\footnotesize
\begin{tabular}{lrrlrrrrrrr} 
\hline\hline
\rule{0pt}{10pt}Loc.     & Abs. gravity & Adj. & Path to mass & Height 	 & Vert.   	& Tie to   & Tie above& Height & Vert.  &Abs. gravity \\
         & at h=130\,cm & for new&				&of tie  	 &Trans.	& NE 	   & magnet   & of mass& trans to & at mass pos. \\
		 & $-9.801\,$m/s$^2$         &grid point&		&		 	&			& marker   &		  & tie    & Mass Pos.& $-9.801\,$m/s$^2$  \\	
		 &	($\mymu$Gal)& ($\mymu$Gal)&				& (cm)		  &($\mymu$Gal)	& ($\mymu$Gal) & ($\mymu$Gal) & (cm) & ($\mymu$Gal) & ($\mymu$Gal) \\[2pt]
\hline
\rule{0pt}{10pt}SE&3114.2(3.2) &2.3(1.2) &SE$\rightarrow$NE$\rightarrow$M &25.9 &260.2(2.1) &22.1(2.6) &-253.9(2.6)& 127.8 &-3.8(3.3) &3141.2(6.3)\\[2pt]
SW&3110.9(3.0)   &1.8(0.9) &SW$\rightarrow$NE$\rightarrow$M &25.9 &257.3(2.1) &25.7(2.3) &-253.9(2.6)& 127.8 &-3.8(3.3) &3138.0(6.1)\\[2pt]
SW&3110.9(3.0) 	&1.8(0.9) &SW$\rightarrow$M, h1&25.9 &257.3(2.1) &         &-227(2.6)   &127.8  &-3.8(3.3) &3139.1(5.0)\\[2pt]
SW&3110.9(3.0) 	&1.8(0.9) &SW$\rightarrow$M, h2&131.1& -2.7(2.0) &            &35.3(2.3)  &128.0  &-3.4(3.1) &3141.9(4.5) \\[2pt]
\hline
\multicolumn{10}{r}{\rule{0pt}{10pt}Weighted mean taking into account the correlations of the four measurements:} & 3140.2(4.3)\\[2pt]
\hline\hline

\end{tabular}
\label{table:ties}
\end{table*}

\section{Error analysis}
The absolute value of $g$ at the mass position for each of the four paths discussed in Section~\ref{sec:tie} is simply the sum of the translations and ties shown in Table~\ref{table:ties}. The error in each value of $g_{m}$, however, is not as simple as a root-sum-square calculation of the individual uncertainties. Since the absolute gravity measurements, as well as the VGG and tie measurements involved the same gravimeter, these values are correlated by systematic uncertainties in the instrument (5\,$\mymu$Gal for the site and instrument uncertainties of the FG5 and 2\,$\mymu$Gal for every VGG and tie). These correlations are accounted for in the analysis. It is best to write the equation in matrix form
\begin{equation}
\mathbf{G} = \mathbf{B}\mathbf{I},
\end{equation}
where $\mathbf{G}$ is a $n$ row single column ($n \times 1$) matrix consisting of the four values of gravity results from the four paths, $g_{i}$, $\mathbf{B}$ is a $n\times m$ matrix containing ones, zeros, and multipliers for $\alpha$ and $\beta$ coefficients, and $\mathbf{I}$ is an $m$ row single column matrix of input data, e.g., tie values and VGG coefficients. The $n\times n$ covariance matrix of $\mathbf{G}$ is then given by
\begin{equation}
\mathbf{V}_\mathrm{G} = \mathbf{B} \mathbf{V}_{\mathrm{I}}\mathbf{B}^\mathrm{T},
\end{equation}
where $\mathbf{V}_I$ is the $m\times m$  covariance matrix of the input data in $\mathbf{I}$. This covariance matrix is sparse and only a few positions outside the diagonal differ from zero. The best estimate of absolute gravity at the mass, $g_{\mathrm{m}}$, can be found using a least-squares analysis of $\mathbf{G}$, accounting for covariance between individual elements.  Following the procedure outlined by Mohr and Taylor~\cite{CODATA98} (adapted from Aitken ~\cite{aitken}), $g_{\mathrm{m}}$ is found such that the squared difference between $\mathbf{G}$ and $g_{\mathrm{m}}$ is minimized while taking into account the correlations. Here, $\mathbf{A}$ is the so-called design matrix, in this case a $4\times 1$ matrix with all elements being unity.

The estimate of $g_{\mathrm{m}}$ which minimizes this squared difference is given by
\begin{equation}
g_{\mathrm{m}} = (\mathbf{A}^\mathrm{T}\mathbf{V}_{\mathrm{G}}^{-1}\mathbf{A})^{-1}\mathbf{A}^\mathrm{T}\mathbf{V}_{\mathrm{G}}^{-1}\mathbf{G}
\end{equation}
and the variance of $g_{\mathrm{m}}$ by
\begin{equation}
\sigma^2_{g_{\mathrm{m}}} = (\mathbf{A}^\mathrm{T}\mathbf{V}_{\mathrm{G}}^{-1}\mathbf{A})^{-1}.
\end{equation}

\section{Results and Future}
\begin{figure}[h]
\centering
\includegraphics[width=3.3in]{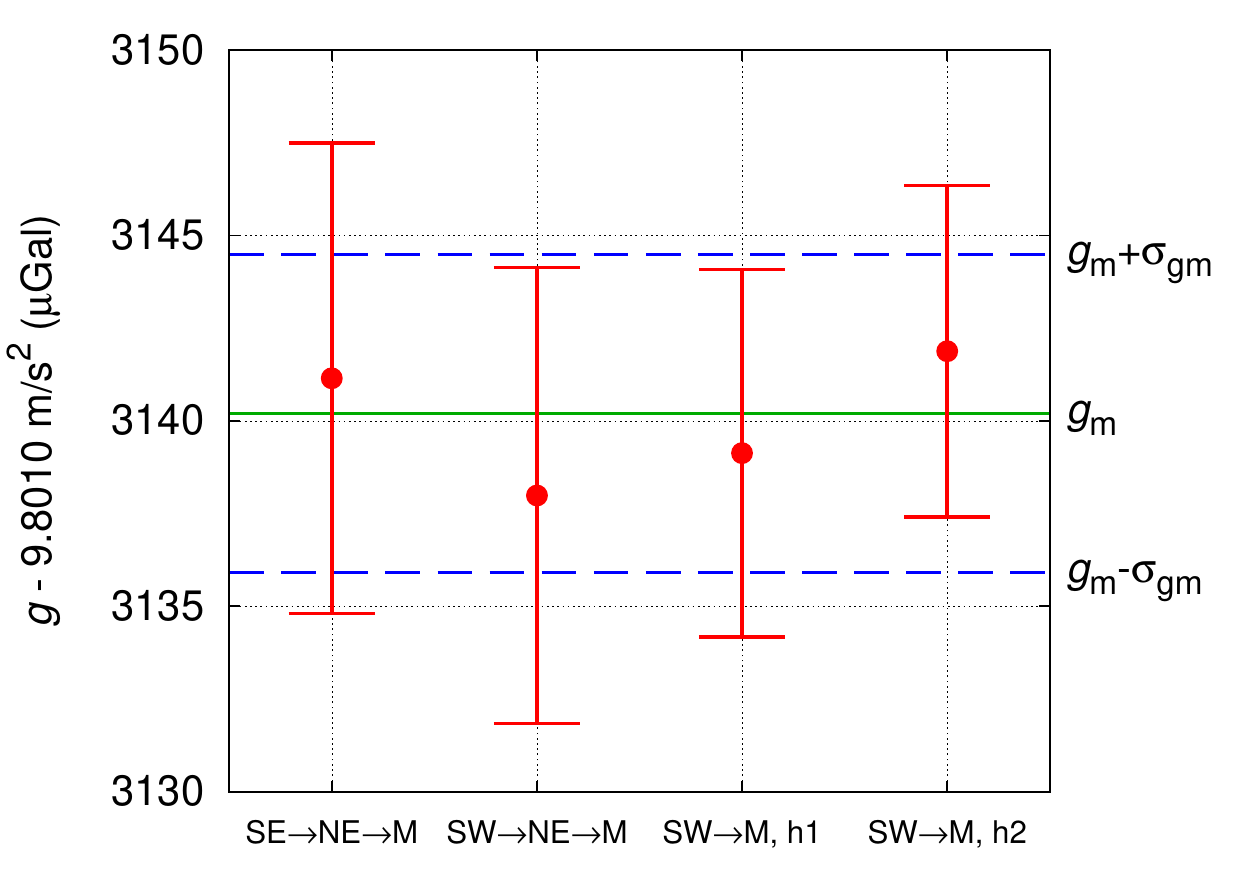}
\caption{Individual absolute gravity determinations at the mass position and the final best estimate.}
\label{fig:results}
\end{figure}

The best estimate analysis gave a final result of 980103140.2(4.3)\,$\mymu$Gal ($4.4\times 10^{-9}$ relative uncertainty). All results are shown in Figure~\ref{fig:results}. These determinations were made almost entirely using measurements, with only a small adjustment made using the model. Originally, the scope of the project did not involve making ties with measurements, but instead relying heavily on the model to determine the gravity transfer between absolute references and the mass position. However, future ties cannot be measured directly above the magnet as NIST-4 construction nears completion. The model now not only serves as a cross-check for measurements, but also as a means to account for any future changes to the laboratory or balance environment. Table~\ref{table:validation} shows the modeled and measured ties between SE and SW at 130\,cm and the mass position. The error in modeled ties with respect to measurements is on the order of only about 1\,$\mymu$Gal or about $10^{-9}$ with respect to $g_{\mathrm{m}}$. 

\begin{table}[h]
\centering
\caption{Modeled and measured ties between absolute reference points at 130\,cm and the mass position.}
\begin{tabular}{p{1.6cm} p{2cm} p{1.8cm} p{1.5cm}} 
\hline
Tie & Measured $\Delta g$ ($\mymu$Gal) & Modeled $\Delta g$ ($\mymu$Gal) & Difference ($\mymu$Gal) \\
\hline
SE \--- M & 26.96 & 26.76 & -0.20 \\
SW \--- M & 28.81 & 30.39 & 1.58 \\
\hline
\end{tabular}
\label{table:validation}
\end{table}

\section{Conclusions}
Gravity variations for the laboratory of the new NIST-4 watt balance were modeled and showed good agreement with a mapping made one year prior. Four (partially) unique ties between absolute gravity reference points and the NIST-4 test mass location were established through relative measurements, and a total uncertainty of only about 4$\times 10^{-9}$ was obtained, well within the necessary range required by the watt balance experiment. Several relatively massive watt balance components were modeled, and modeled ties were shown to be accurate within about 0.7\,$\%$ of one measured tie and within about 5.5\,$\%$ of the other measured tie. In the future, both the model and the presented data will be used for determination of absolute gravity at the precise test mass location once NIST-4 is fully operational.

\section*{Acknowledgments}
The authors would like to thank Jacques Liard for guidance on modeling and measuring gravity.


\begin{thebibliography}{1}
\bibitem{CGPM} Resolution 1 of the 24th Meeting of the General Conference on Weights and Measures (CGPM): On the possible future revision of the International System of Units, Bureau International des Poids et Mesures, Paris, 2011


\bibitem{nist3} S.~Schlamminger, D.~Haddad, F.~Seifert, L.S.~Chao, D.B.~Newell, R.~Liu, R.L.~Steiner and J.R.~Pratt, "Determination of the Planck constant using a watt balance with a superconducting magnet system at the National Institute of Standards and Technology", {\it Metrologia}, {\bf 51}, pp. S15-S24, March 2014.


\bibitem{nist4} D.~Haddad, L.S.~Chao, F.~Seifert, D.B.~Newell, J.R.~Pratt, S.~Schlamminger, “Construction of a watt balance with the aim to realize the kilogram at the National Institute of Standards and Technology,”  {\it Proc. CPEM. Dig.,}  pp. 708-709, August 2014.

\bibitem{nist4grav} D.B.~Newell, J.O.~Liard, L.S.~Chao, A.~Cao, F.~Seifert, D.~Haddad, J.R.~Pratt, and S.~Schlamminger, “The Measurement of the Local Acceleration of Gravity for the NIST-4
Watt Balance,”   {\it Proc. CPEM. Dig.,} pp. 362-363, August 2014.


\bibitem{lnegrav} S.~Merlet, A.~Kopaev, M.~Diament, G.~Geneves, A.~Landragin, and F.~Pereira~Dos~Santos, “Micro-gravity investigations for the LNE watt balance project,” {\it Metrologia} {\bf45}(3), pp. 265-274, April 2008.

\bibitem{metasgrav} H.~Baumann, E.E.~Klingel\'{e}, A.L.~Eichenberger, P.~Richard, and B.~Jeckelmann, “Evaluation of the local value of the Earth gravity field in the context of the new definition of the kilogram,” {\it Metrologia} {\bf 46}(3), pp. 178-186, February 2009.

\bibitem{icac} Z.~Jiang, V.~P\'{a}link\'{a}\v{s}, O.~Francis, P.~Jousset, J.~M\"akinen, S.~Merlet, M.~Becker, A.~Coulomb, K.U.~Kessler-Schulz, H.R.~Schulz, C.~Rothleitner, L.~Tisserand, and D.~Lequin, “Relative Gravity Measurement Campaign during the 8th International Comparison of Absolute Gravimeters (2009),” {\it Metrologia} {\bf 49}, pp.  95-107, December 2011.

\bibitem{NRCgrav} J.O.~Liard, C.A.~Sanchez, B.M.~Wood, A.D~Inglis, and R.J.~Silliker, "Gravimetry for watt balance measurements," {\it Metrologia} {\bf 51}(2), pp. S32-S41, March 2014.

\bibitem{nist3grav} D.B.~Newell, J.O.~Liard, A.D~Inglis, M.C.~Eckl, D.~Winester, R.J.~Silliker, and C.G.L.~Gagnon, “The possible contribution of gravity measurements to the difference between the NIST and NRC watt balance results,” {\it Metrologia} {\bf 50}(4), pp. 337-344  June 2013, E50(5), 557-558,  September 2013.


\bibitem{sibro} "The International System of Units (SI)", 8th edn (S\`{e}vres, France: Bureau International des Poids et Mesures) table 9, www.bipm.org/en/si/si brochure/chapter4/table9.html

\bibitem{nagy66} D.~Nagy, "The gravitational attraction of a right rectangular prism," {\it Geophysics} {\bf 31}(2), pp. 362-371, April 1966.


\bibitem{scintrex} Scintrex Limited, 222 Snidercroft Road, Concord L4K 2K1 Ontario, Canada.

\bibitem{microg} Micro-g LaCoste, 1401 Horizon Ave., Lafayette, CO 80026.

\bibitem{CODATA98} P.J.~Mohr, B.N.~Taylor, "CODATA recommended values of the fundamental physical constants: 1998," {\it Journal of Physical and Chemical Reference Data}, {\bf 28}(6), pp. 1713-1852, April 2000.

\bibitem{aitken} A.C.~Aitken, 1934, {\it Proc. R. Soc. Edinburgh} {\bf 55}, pp. 42–48, 1934.

\end{thebibliography}
\end{document}